\documentclass[prc,twocolumn,superscriptaddress,showpacs,amssymb,amsmath,amsfonts,aps]{revtex4}
\usepackage{graphicx}
\usepackage{dcolumn}
\usepackage{epsfig}
\begin{document}
\title{Electroexcitation of the Roper resonance\footnotetext{Notice: Authored by
The Southeastern Universities Research Association, Inc. under U.S. DOE Contract
No. DE-AC05-84150. The U.S. Government retains a non-exclusive, paid-up, irrevocable,
world-wide license to publish or reproduce this manuscript for U.S. Government purposes.
}\\in relativistic quark models}

\newcommand*{\JLAB }{ Thomas Jefferson National Accelerator Facility, Newport
News, Virginia 23606, USA}
\affiliation{\JLAB }

\newcommand*{\YEREVAN }{ Yerevan Physics Institute, 375036 Yerevan, Armenia}
\affiliation{\YEREVAN }

\author{I.G.~Aznauryan}
     \affiliation{\JLAB}\affiliation{\YEREVAN}

\thispagestyle{empty}

\begin{abstract}
{The amplitudes of the transition
$\gamma^* N\rightarrow P_{11}(1440)$ are calculated within
light-front relativistic quark model
assuming that the $P_{11}(1440)$
is the first radial excitation
of the $3q$ ground state.
The results are presented along with the predictions obtained in 
other relativistic quark models.
In comparison with the previous calculations, we have extended
the range of $Q^2$ up to $4.5~GeV^2$
to cover
the kinematic interval of the forthcoming experimental data.
Using approach based on PCAC,
we have checked the relative sign between
quark model predictions for the $N$ and $P_{11}(1440)$ contributions 
to the $\pi$ electroproduction found
in previous investigations. 
}
\end{abstract}

\maketitle

\section{Introduction}
The amplitudes of the electroexcitation
of the Roper resonance on proton  
are expected to be obtained from the CLAS 
$\pi^+$ electroproduction data
at $1.7<Q^2<4.2~GeV^2$. There are already
results \cite{1} extracted from 
the preliminary data \cite{2,3}, and the final results
will be available soon.
The results at $1.7<Q^2<4.2~GeV^2$
combined with the previous CLAS
data at $Q^2=0.4,~0.65~GeV^2$ \cite{4,5,6}
and with the information at $Q^2=0$ \cite{7} will
give us knowledge of 
the $Q^2$ evolution of the $P_{11}(1440)$ 
electroexcitation
in wide $Q^2$ region. This information can be very important
for understanding of the nature of the Roper resonance
which has been the subject 
of discussions since its
discovery \cite{8}, because the simplest and most natural
assumption that this is a first radial excitation
of the $3q$ ground state led to the difficulties
in the description of the mass of the resonance.

To use the  information
on the the electroexcitation
of the Roper resonance,
it is important to 
understand 
the quark model predictions for 
the $Q^2$ evolution of the transition
$\gamma^* N\rightarrow P_{11}(1440)$.
It is known that
with increasing $Q^2$, when the momentum
transfer becomes larger than the masses of the constituent quarks,
a relativistic treatment of the electromagnetic excitations,
which is important already at $Q^2=0$,
becomes crucial.
The consistent way to realize the relativistic
treatment of the
$\gamma^* N \rightarrow N^*$  transitions 
is to consider
them in the light-front (LF) dynamics \cite{9,10,11}.
Within this framework, one can set up an impulse approximation
and avoid difficulties caused by different
momenta of initial and final hadrons 
in the $\gamma^* N \rightarrow N^*$  transition.

For the
$P_{11}(1440)$, the LF calculations have been realized 
in Refs. \cite{12,13,14,15,16}.
However, 
quantitatively 
the obtained results 
differ from each other and there are inconsistencies 
in the signs of the amplitudes
presented in Refs. \cite{12,13,14,15,16}.
In particular, 
the relative sign between the $NN\pi$ and
$P_{11}(1440)N\pi$ amplitudes, 
which is necessary for comparison of quark-model
predictions for the $\gamma^* N\rightarrow P_{11}(1440)$
amplitudes with experimental data, was found and taken
into account only 
in Ref. \cite{13}, where
for this purpose
the ${}^3P_0$ model of Ref. \cite{17} was used.

For this reason, we found it important
to perform calculations of the 
$\gamma^* N\rightarrow P_{11}(1440)$ amplitudes
in the LF relativistic quark model, and also to check
the relative sign between the $NN\pi$ and
$P_{11}(1440)N\pi$ amplitudes,
using a different approach. We will present and discuss
the results of all relativistic quark models.
All results will be given within a unified
definition of the $\gamma^* N\rightarrow P_{11}(1440)$
helicity amplitudes.

We will find the relative sign between the $NN\pi$ and
$P_{11}(1440)N\pi$ amplitudes
by using the hypothesis of
partially conserved axial-vector current (PCAC)
in the way used in Ref. \cite {18}
for the description of the $hadron\rightarrow hadron+\pi$
transitions. This method was previously used in Ref. \cite {19}
for evaluation of the relative signs between the $NN\pi$
and $N^*N\pi$ amplitudes for the resonances of the multiplet
$[70,1^-]$. 
Using an approach based on PCAC
we have confirmed the sign obtained in Ref. \cite {13}.

In Sec. II we will
specify definitions
of the
$\gamma^* p\rightarrow P_{11}(1440)$ 
helicity amplitudes
in quark models and will give relations between  
these amplitudes and the amplitudes extracted from 
experimental data. 
In Sec. III the formulas for the
$\gamma^* p\rightarrow P_{11}(1440)$ amplitudes
obtained
in the LF relativistic quark model of Ref. \cite {20} 
will be given. The relative sign between the $NN\pi$ and
$P_{11}(1440)N\pi$ amplitudes
will be found in Sec. IV.
The results of numerical calculations
of the 
$\gamma^* p\rightarrow P_{11}(1440)$ amplitudes
will be presented and discussed in Sec. V along with the
results obtained in Refs. \cite{12,13,14,15,16}.
Summary will be done in Sec. VI.

\section{Definitions of the 
$\gamma^* N\rightarrow P_{11}(1440)$ helicity amplitudes}
The relations between
the $\gamma^* p\rightarrow P_{11}(1440)$ helicity amplitudes
extracted from $\pi$ electroproduction data
and the $P_{11}(1440)$ contribution
to the   
$\gamma^* p\rightarrow 
\pi N$ multipole amplitudes
have the following form
\cite{21,22}:
\begin{eqnarray}
&&A_{1/2}=aImM^R_{1-}(W=M),\\
&&S_{1/2}=-\frac{a}{\sqrt{2}}ImS^R_{1-}(W=M),\\
&&a=
\sqrt{2\pi\frac{q_{\pi}}{K}\frac{M}{m}
\frac{\Gamma}{\beta_{\pi N}}}/C,\\
&&C=-\sqrt{\frac{1}{3}}~~for~~\gamma^* p\rightarrow p\pi^0,\\
&&C=-\sqrt{\frac{2}{3}}~~for~~\gamma^* p\rightarrow n\pi^+.
\end{eqnarray}
Here $\Gamma$ and $M$ are the total width and mass of the 
resonance, $\beta_{\pi N}$ is its branching ratio
to $\pi N$ channel,
$m$ is the mass of the
nucleon, $K\equiv (M^2-m^2)/2M$, $q_{\pi}$ is the center of mass 
momentum
of the pion at the resonance position.

The helicity amplitudes defined through Eqs. (1-5)
include the relative sign between $g_{NN\pi}$ and
$g_{RN\pi}$ coupling constants that determine the relative
contributions of
the Born term and the $P_{11}(1440)$
to the reaction $\gamma^* N\rightarrow\pi N$.

In Refs. \cite{12,14,15,16} and in this work,
the quark model predictions for the transition
$\gamma^* N\rightarrow P_{11}(1440)$ are found in terms
of form factors that enter
the matrix element of the transition current.
Following Ref. \cite{12}, we define this matrix element in 
the following form:
\begin{eqnarray}
&&<N^*|J_{\mu}|N>=e\bar{u}(P^*)\Gamma_{\mu}u(P),\\
&&\Gamma_{\mu}=
\left(q^2\gamma_{\mu}-(q\gamma)q_{\mu}
\right)F^*_1(Q^2)
+i\sigma_{\mu\nu}q^{\nu}F^*_2(Q^2).
\end{eqnarray}

Helicity amplitudes of the $\gamma^* p\rightarrow P_{11}(1440)$ 
transition are related to
the form factors $F^*_{1p}(Q^2)$ and $F^*_{2p}(Q^2)$
by:
\begin{eqnarray}
&A^q_{1/2}=c\left[-Q^2F^*_{1p}(Q^2)+(M+m)F^*_{2p}(Q^2)\right] , \\
&S^q_{1/2}=-c\frac{q^{*}_{cms}}{\sqrt{2}}
\left[(M+m)F^*_{1p}(Q^2)+F^*_{2p}(Q^2)\right],\\
&c=\sqrt{\pi\alpha\frac{Q^2+(M-m)^2}{MmK}},\nonumber
\end{eqnarray}
where $\alpha=e^2/4\pi=1/137$.

The expressions for the helicity amplitudes
$A^q_{1/2},~S^q_{1/2}$ in Refs. \cite{14,15,16}
coincide with Eqs. (8,9).
In Ref. \cite{12}, the formulas for both
amplitudes contain inaccuracies in the coefficients
that we have corrected in the results from that work
presented below.
 
In quark model calculations another definition of
the $\gamma^* N\rightarrow P_{11}(1440)$
helicity amplitudes 
through the $\gamma^* N\rightarrow N^{*}$
transition current matrix elements 
also is used \cite{12,13,23}:
\begin{eqnarray}
&&A^q_{1/2}=b
<N^{*+},S_z^*=\frac{1}{2}|J\epsilon|p,S_z=-\frac{1}{2}>,\\
&&S^q_{1/2}=b
\frac{q^*_{cms}}{Q}
<N^{*+},S_z^*=\frac{1}{2}|J\epsilon|p,S_z=\frac{1}{2}>,\\
&&b=\left[\frac{2\pi 
\alpha}{K}\right]^{1/2} \nonumber.
\end{eqnarray}
Here it is supposed that the virtual photon moves
along the z-axis in the $N^*$ rest frame and its 3-momentum is
$q^*_{cms}$, $P^*=P+q$, $Q^2\equiv -q^2$.

The helicity amplitudes $A^q_{1/2},~S^q_{1/2}$
found in different works can not be directly compared
with each other and with amplitudes extracted from
experimental data for the following reasons: 

(i) The signs of the amplitudes $A^q_{1/2},~S^q_{1/2}$
in quark models depend on the sign
of the $P_{11}(1440)$ wave function,
which is different in different works.
Let us define 
the $P_{11}(1440)$ wave function 
in the nonrelativistic approximation in the form:
\begin{equation}
\Psi_R \sim \zeta (\sum {\bf{k}}_i^2
-\alpha^2)\Psi_N,
\end{equation}
where ${\bf{k}}_i~(i=1,2,3)$ are the 
quark
three-momenta in the center-of-mass system and $\alpha$ is 
the parameter
that depends on interquark forces.
With this definition,
the wave functions of Refs. \cite{12,14,15,16}
and  Ref. \cite{13} correspond
to $\zeta=+$ and $-$, respectively.
Our definition of 
the sign of the $P_{11}(1440)$ wave function used below 
coincides with that
in Ref. \cite{13}. 

(ii) The amplitudes $A^q_{1/2}$, $S^q_{1/2}$
found in quark models through relations (8-11)
are related to the amplitudes $A_{1/2}$, $S_{1/2}$
extracted from experimental data using definitions (1-5)
in the following way
\begin{equation}
A_{1/2}=-\xi_RA^q_{1/2},~ S_{1/2}=-\xi_RS^q_{1/2}.
\end{equation}
Here $\xi_R$ is 
the relative sign between $g_{NN\pi}$ and
$g_{RN\pi}$ coupling constants, which also depends on $\zeta$.
The sign $\zeta$
drops out from the amplitudes $A_{1/2}$, $S_{1/2}$,
therefore, these amplitudes do not depend on
the sign of the $P_{11}(1440)$ wave function.

In the previous investigations, the sign $\xi_R$
has been found and taken into account only in Ref. \cite{13},
where it was obtained
that $\xi_R=-\zeta$.
We have obtained the same sign in this work using a different
approach.

In Refs. \cite{12,14,15,16}, only the results for the 
amplitudes $A^q_{1/2}$, $S^q_{1/2}$ are given.
Corresponding results
for the amplitudes $A_{1/2}$, $S_{1/2}$ 
we will present using the sign $\xi_R$
found in this work and in Ref. \cite{13}.

\section{The 
$\gamma^* p\rightarrow P_{11}(1440)$ amplitudes in the
relativistic quark model}

The calculations of the
$\gamma^* N\rightarrow P_{11}(1440)$ amplitudes
we have performed
in the relativistic quark model of Ref. \cite{20},
constructed for radiative transitions
of hadrons in the infinite momentum frame (IMF),
where  
\begin{eqnarray}
&& \mathbf{P}\parallel z,~~P\rightarrow \infty, \\  
&& q=\left(\mathbf{q}_{\perp}, -\frac {M^2-m^2-\mathbf{q}^2_{\perp}}{4P},
\frac {M^2-m^2-\mathbf{q}^2_{\perp}}{4P}\right). \nonumber
\end{eqnarray}
Here $P$ is the momentum of the initial hadron and $q$ is
the photon momentum, $Q^2=\mathbf{q}_{\perp}^2$. 
Such approach is analogous to the
LF calculations \cite{9,10,11,12,13,14,15,16}. 
The derivation of the formulas for the transition
$\gamma^* N(\frac{1}{2}^+)\rightarrow N(\frac{1}{2}^+)$
are presented in detail in Ref. \cite{24}, 
where the model
of Ref. \cite{20} was used for investigation of the $Q^2$
evolution of the nucleon
form factors. Here we will give 
final expressions for the form factors $F^*_{1p}(Q^2)$,
$F^*_{2p}(Q^2)$ that follow from the results
of Ref. \cite{24}:
\begin{eqnarray}
&&Q^2F^*_{1p}(Q^2)=\int {\cal 
F}_{1p}\Phi_N(M^2_0)\Phi_R(M'^2_0)d\Gamma,\\ 
&&QF^*_{2p}(Q^2)=\int {\cal F}_{2p}\Phi_N(M^2_0)\Phi_R(M'^2_0)d\Gamma, 
\end{eqnarray}
where $\Phi_N(M^2_0)$ and $\Phi_R(M'^2_0)$ are radial parts
of the nucleon and $P_{11}(1440)$ wave functions,
$M_0$ and $M'_0$ are invariant masses of quarks in the
initial and final hadrons.

Let us parametrize quark momenta in the initial
and final hadrons by:
\begin{equation}
\mathbf{p}_i=x_i\mathbf{P}+\mathbf{k}_{i\perp},~~
\mathbf{p}'_i=x_i\mathbf{P}'+\mathbf{k}'_{i\perp},
\end{equation}
where $i=a,b,c$ denotes the quarks in the hadrons,
and we will suppose that the current interacts
with the quark $a$. It is convenient to
parametrize the variables $x_i,\mathbf{k}_{i\perp}$
and $\mathbf{k}'_{i\perp}$ in the following way:
\begin{eqnarray}
&&\mathbf{k}_{i\perp}=\mathbf{K}_{i\perp}-\frac{1}{2}y_iQ,
~~\mathbf{k}'_{i\perp}=\mathbf{K}_{i\perp}+\frac{1}{2}y_iQ,\\  \nonumber
&&x_a=1-\eta,~~y_a=x_a-1,~~\mathbf{K}_{a\perp}=-\mathbf{K}_{\perp},\\\nonumber
&&x_b=(1-\xi)\eta,~y_b=x_b,~
\mathbf{K}_{b\perp}=-\mathbf{k}_{\perp}+(1-\xi)\mathbf{K}_{\perp},\\ \nonumber 
&&x_c=\xi\eta,~~y_c=x_c,~~ 
\mathbf{K}_{c\perp}=\mathbf{k}_{\perp}+\xi\mathbf{K}_{\perp}. \nonumber
\end{eqnarray}
In terms of the variables 
$\mathbf{k}_{\perp},\mathbf{K}_{\perp},\xi,\eta$
the phase space volume has the form
\begin{equation}
4(2\pi)^6d\Gamma=\frac
{d\mathbf{k}_{\perp}d\mathbf{K}_{\perp}d\xi d\eta}
{\xi(1-\xi)\eta(1-\eta)},
\end{equation}
the invariant masses of quarks are equal to
\begin{eqnarray}
&&M_0^2(M'^2_0)=\frac{\mathbf{K}_{\perp}^2+m_q^2\eta}{\eta(1-\eta)}
+\frac{\mathbf{k}_{\perp}^2+m_q^2}{\eta\xi(1-\xi)}\\
&&+\frac{\eta Q^2}{4(1-\eta)}\mp\frac{K_x Q}{1-\eta},\nonumber
\end{eqnarray}
and the functions ${\cal F}_{1p},~{\cal F}_{2p}$
in Eqs. (15) and (16) have the form
\begin{eqnarray}
&&{\cal F}_{1p}=\frac{R^a_{++}(R^b_{++}R^c_{--}-R^b_{+-}R^c_{-+})}
{D_aD_bD_c},\\
&&{\cal F}_{2p}=\frac{R^a_{-+}(R^b_{-+}R^c_{+-}-R^b_{--}R^c_{++})}
{D_aD_bD_c},
\end{eqnarray}
where
\begin{eqnarray}
&&R^i_{\pm \pm}=m_i m'_i+\mathbf{K}^2_{i\perp}-
\frac{1}{4}y_i^2Q^2\pm iy_iQK_{iy},\\
&&R^i_{\pm \mp}=\mp \frac{1}{2}y_i(m_i+m'_i)Q\\ \nonumber
&&~~~~~~~~+x_i(M_0-M'_0)(iK_{iy}\mp K_{ix}),\\ 
&& m_i=m_q+M_0 x_i,~~m'_i=m_q+M'_0 x_i,\\
&&
D_i=[(m_i^2+\mathbf{k}^2_{i\perp})(m_i^2+\mathbf{k}'^2_{i\perp})]^{1/2}, 
\end{eqnarray}
$m_q$ is the quark mass,
$i=a,b,c$. 

Under the assumption
that $P_{11}(1440)$ is a radial excitation of the
nucleon considered as the $3q$
ground state, we have defined $\Phi_R(M^2_0)$
in the form: 
\begin{equation}
\Phi_R(M^2_0)=N(\beta^2-M_0^2)\Phi_N(M^2_0).
\end{equation}
The parameters $N$ and $\beta$ are determined by the conditions:  
\begin{equation}
\int \Phi_R(M^2_0)\Phi_N(M^2_0)d\Gamma=0,~~\int 
\Phi^2_{N(R)}(M^2_0)d\Gamma=1. 
\end{equation}

As in Refs. \cite{20,24}, we will take
the radial part of the nucleon wave function 
in the Gaussian form:
$\Phi_N(M_0^2)\sim exp(-M_0^2/6\alpha_{HO}^2)$. 
The only parameters of the approach, the quark mass 
($m_q=0.22~GeV$) and 
the harmonic-oscillator parameter ($\alpha_{HO}=0.38~GeV$),
were found in Ref. \cite{20} from the description
of the static properties of the nucleon.

\section{The relative sign between the $NN\pi$ and 
$P_{11}(1440)N\pi$ amplitudes}

In order to find the relative sign between the $NN\pi$ and
$P_{11}(1440)N\pi$ amplitudes, we 
utilize PCAC
in the way used in Ref. \cite {18}
for the description of the $hadron\rightarrow hadron+\pi$ 
transitions.
PCAC relates the divergence of the charged axial-vector 
current to the pion field:
\begin{equation}
\partial_
{\lambda}(J^{a\lambda}_1+iJ^{a,\lambda}_2)=
m^2_{\pi}f_{\pi}\phi_{\pi^-},
\end{equation}
where $f_{\pi}=132~MeV$ is the $\pi\rightarrow \mu\nu$
coupling constant.

In IMF (14), 
the matrix element of Eq. (29) 
between two
spin $\frac{1}{2}$ particles with momenta $P_1$ and $P_2$
at $\mathbf{q}_{\perp}=0$ 
gives:
\begin{equation}
f_{\pi}g(N_1N_2\pi)=
g^{12}_A\frac{m_1+m_2}{\sqrt{2}},
\end{equation}
where
\begin{equation}
g^{12}_A=\frac{1}{2P} 
<P_2,S_z=\frac{1}{2}|J^{a,0}_{1+i2}|
P_1,S_z=\frac{1}{2}>|_{P\rightarrow \infty}
\end{equation}
and $g(N_1N_2\pi)$ is the coupling constant
in the $N_1N_2\pi$ vertex:
\begin{equation}
{\cal L}_{N_1N_2\pi}=ig(N_1N_2\pi)\bar{\psi_2}\gamma_5 
{\vec{\tau}}\psi_1{\vec{\phi}}.
\end{equation}
According to the results of Ref. \cite{20},
we have:
\begin{eqnarray}
&g^{12}_A=\frac{5}{3}\int 
\frac{(m_q+M_0x_a)^2-\mathbf{K}^2_{\perp}}
{(m_q+M_0x_a)^2+\mathbf{K}^2_{\perp}}
\Phi_1(M^2_0)\Phi_2(M^2_0)d\Gamma. 
\end{eqnarray}
Using Eqs. (30) and (33) one can find the relative sign
between  the $NN\pi$ and
$P_{11}(1440)N\pi$ coupling constants: $\xi_{R}$.
In the limit when $\alpha_{HO}\ll m_q$,
the ratio of the coupling constants (31)
for the $P_{11}(1440)N$ and $NN$ transitions
that follows from Eq. (33) can be found
in analytical form and is equal to: 
\begin{equation}
\frac{g_A^{RN}}{g_A^{NN}}=-\zeta\frac{\alpha^2_{HO}}{3\sqrt{6}m_q^2}.
\end{equation}
We have also calculated the ratio (34) numerically
for the values of the quark mass in the range $0<m_q<m/3$
and
in the wide range of the harmonic oscillator
parameter which includes the values of  
$\alpha_{HO}$ 
from Refs. \cite{12,13,20}. 
For all these values of $m_q$ and $\alpha_{HO}$,
we have found the same sign as in Eq. (34).
So, using approach based on PCAC we have obtained
that the relative
sign between $NN\pi$ and $P_{11}(1440)N\pi$
coupling constants is equal to $\xi_{R}=-\zeta$. 
The same sign was obtained in Ref. \cite{13}
using ${}^3P_0$ model.
\section{Results}
Our results for the $\gamma^* p \rightarrow P_{11}(1440)$
helicity amplitudes are presented in Fig. 1 along with the results
obtained in Refs. \cite{12,13,14,15,16}
and with experimental data. 
The thick lines correspond to the results obtained
in the LF formalism
using the $\it {plus}$ component of the electromagnetic
current $J^+\equiv J^0+J^3$ in the frame where
$q^+\equiv q^0+q^3=0$. These results are more favorable,
because in this case the contributions of
virtual $q\bar{q}$ pairs are eliminated, and the processes
$\gamma^* N\rightarrow N^*$ are determined only
by the $N(N^*)\rightarrow 3q$ vertices 
(Refs. \cite{9,10,11,20,25}). 
Therefore, the matrix elements
of $J^+$ in the frame $q^+=0$ can be presented in analogy
with the nonrelativistic case through a sum of one-body
currents for constituent quarks.  
Let us stress that in such approach when we 
evaluate electromagnetic form factors by utilizing only
$\it {plus}$ component of the electromagnetic
current, it is supposed that other components 
have proper behavior in order to fulfill
the current conservation.

The thin dash-dotted lines are the results of Ref. \cite{15}
obtained in the LF formalism; however, in this investigation
transverse component of the electromagnetic current,
which can contain contributions of virtual $q\bar{q}$ pairs,
is also used.

The thin solid and dashed lines correspond to the results
of Ref. \cite{16} obtained in the relativistic quark models
with instant-form and point-form kinematics.

The results of all quark models presented in Fig. 1
correspond to the sign $\xi_{R}$
found in this work and in Ref. \cite{13}.  

It is important to note that in all approaches
\cite{12,13,14,15,16} and in the approach
used in this work, a good description of elastic 
nucleon form factors has been obtained.
Within the approach of this work,
the proton and neutron electromagnetic form factors
and the nucleon axial-vector form factor 
were described up to $Q^2=4~GeV^2$ \cite{24}.
The predictions obtained in Ref. \cite{24} are also in reasonable 
agreement
with the experimental data obtained later, for example,
with the data from Refs. \cite{26,27}.
A good description of 
nucleon form factors up to $Q^2=4~GeV^2$ was obtained
also in Ref. \cite{13}.
In the approaches of Refs. \cite{14,15,16}
the available data on nucleon form factors were described
up to $Q^2=10~GeV^2$. The approach of Ref. \cite{12}
describes these form factors in more narrow 
region of $Q^2$: $Q^2<1.5~GeV^2$.

In this work and in Refs. \cite{12,13,16},
a simple algebraic form for the radial part of the $3q$ ground
state wave function is used. In our approach
and in Refs. \cite{12,13},
this is a Gaussian form with one parameter:
the harmonic-oscillator parameter. In 
Ref. \cite{16}, the radial part of the ground
state wave function has powerlike form with two parameters. 
The parameters of all these approaches, except Ref. \cite{13}, 
were found from the description of elastic nucleon
form factors in the ranges of $Q^2$ discussed above.
In Ref. \cite{13}, the harmonic-oscillator parameter
was taken from Ref. \cite{28}, were it was found from the description
of baryon masses in the nonrelativistic approximation.
This parameter is very close to that used in this
work and found in Refs. \cite{20,24} from the description
of the nucleon elastic form factors.

In contrast with the approaches used in this work and in Refs. 
\cite{12,13,16}, the wave functions of Refs. \cite{14,15}
were found from the solution of the
equation for the light-front mass operator with
the one-gluon exchange between quarks taken in the form 
used in the relativized quark model of Ref. \cite{29}. 
However, it turned out that for the description
of the nucleon form factors with the nucleon wave function
found in this way, a significant deviation
from the pointlike constituent quarks should be introduced.
Let us emphasize that in Refs. \cite{12,13,16,24}, the 
nucleon form factors were described
with pointlike constituent quarks.
Another difference with the wave functions used
in this work and in Refs. \cite{12,13,16} is the presence
of the configuration mixings in the $N$ and $P_{11}(1440)$
wave functions that arise from the 
one-gluon exchange between quarks. 

The values of the quark mass 
used in this work
and in  Refs. 
\cite{13,14,15,16} are close to
the light-quark mass obtained from
the description of baryon masses in Ref. \cite{29}.

From Fig. 1, it can be seen that 
although all approaches under consideration
give good description
of nucleon form factors,
the predictions for the
$\gamma^* p \rightarrow P_{11}(1440)$ helicity amplitudes
are different. As it was mentioned above,
we consider as more preferable the predictions obtained in
the LF approaches based on the utilization
of the $\it{plus}$ component of the electromagnetic current.
These approaches 
are close to each other,
and the differences in their predictions
are caused mainly by the 
differences
in the N and $P_{11}(1440)$ wave functions.
In spite of the differences, 
the predictions obtained in the LF approaches
have common features that are in agreement
with existing experimental data:
\begin{itemize}
\item The sign of the transverse amplitude
$A_{1/2}$ at $Q^2=0$ 
is negative.
\item The sign
of the longitudinal
amplitude $S_{1/2}$ at small $Q^2$ is positive.
\item All LF approaches predict the change of the sign
of $A_{1/2}$ at small $Q^2$. The existing experimental data
show a tendency that is in agreement with such sign change.
The confirmation of this prediction
by forthcoming experimental data will be important
for understanding of the nature of the $P_{11}(1440)$
resonance.
\end{itemize}
These predictions are obtained
assuming that the $N$ and the $P_{11}(1440)$
resonance are the $3q$ ground state
and the first radial excitation
of the $3q$ ground state, respectively. The admixtures of other
configuration states in the $N$ and $P_{11}(1440)$
taken into account
in Refs. \cite{14,15} do not affect these results.

\begin{figure*}[tb]
\begin{center}
\epsfig{file=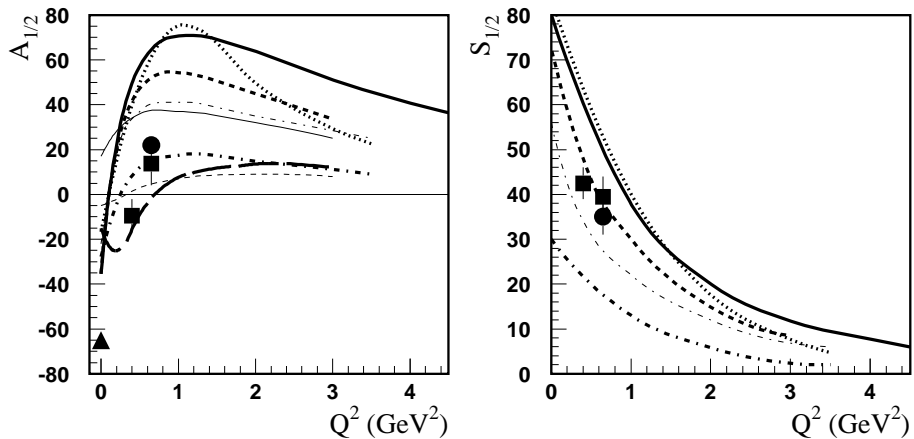,width=12cm}
\end{center}
\caption{Helicity amplitudes
for the  $\gamma^* p \rightarrow P_{11}(1440)$ transition
(in $10^{-3}~GeV^{-1/2}$ units).
Thick solid lines are the results obtained in this work.
Thick dashed, dash-dotted and
long-dashed lines 
correspond to the LF
relativistic quark models
of Refs. \cite{13,14,16}, respectively.
Thick dotted lines correspond to the results of Ref. \cite{12}
corrected according to the discussion in Sec. II.
Thin dash-dotted lines are the results of Ref. \cite{15}.
Thin solid and dashed lines correspond to
the results of relativistic quark models 
with instant-form and point-form kinematics of Ref. \cite{16}.
Full  boxes and circles are  the results obtained
in the analysis of $\pi$ electroproduction data in Ref. \cite{4}
and in the combined analysis of $\pi$ and $2\pi$ 
electroproduction data in Ref. \cite{5}, respectively.
Full triangle at $Q^2=0$ is the 
PDG estimate \cite{7}.} 
\label{fig:fig1}
\end{figure*}
\section{Summary}
In this work we have calculated the 
$\gamma^* p\rightarrow P_{11}(1440)$ transition amplitudes in the
LF relativistic quark model in the range of $Q^2$:
$0\leq Q^2 \leq 4.5~GeV^2$.

We have also calculated the relative sign
between $g_{NN\pi}$ and $g_{RN\pi}$ coupling constants
which is necessary for comparison
of quark-model predictions for the
$\gamma^* p\rightarrow P_{11}(1440)$ amplitudes
with those extracted from experimental data.
The relative sign between the $NN\pi$ and
$P_{11}(1440)N\pi$ vertices was found
by relating these vertices to
the matrix elements of the axial-vector current
using PCAC.
The obtained sign is in agreement with the sign
found in Ref. \cite{13} using different approach,
namely, the ${}^3P_0$ model of Ref. \cite{17}.

The obtained $P_{11}(1440)$ electroexcitation
helicity amplitudes are presented along with the
results found in relativistic quark
models of Refs. \cite{12,13,14,15,16}.
All results are given 
using the sign $\xi_R$ found in this work
and in Ref. \cite{13}. 

All approaches under consideration
give good description
of nucleon form factors; however
the predictions for the 
$\gamma^* p \rightarrow P_{11}(1440)$ helicity amplitudes
are different. 

In Fig. 1, the thick lines present the results
obtained in the LF approaches using the
$\it {plus}$ component of the electromagnetic
current, which we consider as more favorable.
These approaches 
are  close to each other, and the differences
in their predictions
are caused mainly by the differences
in the $N$ and $P_{11}(1440)$ wave functions.
Nevertheless, 
there are several features which are common
for the predictions 
of all LF approaches.
One of the features is the nontrivial
behavior of the transverse amplitude $A_{1/2}$
connected with the change of the sign 
of $A_{1/2}$ with increasing $Q^2$.
The existing experimental data cover a
very limited region of small $Q^2$ up to $0.65~GeV^2$ and
have a tendency that is in agreement with such behavior.
The confirmation of this prediction
by forthcoming experimental data 
will be very important
for understanding of the nature of the $P_{11}(1440)$
resonance.

All approaches fail to describe the  value 
of the transverse amplitude $A_{1/2}$ at $Q^2=0$. 
This can be an indication
on the large pion cloud contribution 
to the $\gamma^* N \rightarrow P_{11}(1440)$ transition
that is expected to be significant
at small $Q^2$.
When the final data on the helicity amplitudes at
$1.7<Q^2<4.2~GeV^2$ will become available, complete 
simultaneous description of the nucleon form factors
and the $\gamma^* N \rightarrow P_{11}(1440)$ amplitudes
will be necessary. Such research will be important
to find the size of the pion cloud contribution,
to specify the form of the $N$ and $P_{11}(1440)$ wave functions, 
including the configuration mixings in these
states, to understand the role of 
the quark form factors and anomalous magnetic moments,
as well the role of other
effects, for instance, the quark mass 
dependence 
on the quark virtualities and therefore on $Q^2$.

\section{Acknowledgments}
I am grateful to the participants of the JLab Theory Group seminar
for interest to this problem that stimulated this work,
and to V.Burkert, H.Lee, A.Thomas for encouraging interest.
I am thankful to B.Julia-Diaz and
S.Stepanyan for their help in numerical calculations,
to T.Sato for discussions of the definitions of
the $\gamma^* N \rightarrow P_{11}(1440)$ helicity amplitudes,
to H.Matevosyan for discussion of the problems
related to the pion cloud contribution,
and to M.Paris for valuable remarks while reading
the draft of this article.

\end{document}